\documentclass{article}
\usepackage[dvips]{graphicx}
\usepackage{amssymb}
\usepackage[cp1251]{inputenc}
\usepackage{mathtext}
\usepackage{caption}
\usepackage{authblk}
\usepackage[left=3cm,right=2cm,top=3cm,bottom=3cm,bindingoffset=0cm]{geometry}

\newcommand {\SPS}    {Sn$_2$P$_2$S$_6$}
\newcommand {\PbPS}    {Pb$_2$P$_2$S$_6$}

\newcommand {\SPbPS}    {(Pb$_y$Sn$_{1-y}$)$_2$P$_2$S$_6$}
\newcommand {\SnPb}    {Sn(Pb)$_2$P$_2$S$_6$}
\begin{document}
\title{Valence fluctuations in \SnPb\ ferroelectrics}
\author[1]{R.~Yevych}
\author[1]{V.~Haborets}
\author[1]{M.~Medulych}
\author[1]{ A.~Molnar}
\author[1]{A.~Kohutych}
\author[2]{A.~Dziaugys}
\author[2]{Ju.~Banys}
\author[1]{Yu.~Vysochanskii}
\affil[1]{Uzhhorod National University, Uzhhorod, Ukraine, 88000}
\affil[2]{Vilnius University, Vilnius, Lithuania, LT-10222}
    \maketitle
    \begin{abstract}
    The valence fluctuations which are related to the charge disproportionation of phosphorous ions $\rm{P}^{4+} + \rm{P}^{4+}\rightarrow\rm{P}^{3+} + \rm{P}^{5+}$ are the origin of ferroelectric and quantum paraelectric states in \SnPb\ semiconductors. They involve recharging of SnPS$_3$ or PbPS$_3$ structural groups which could be represented as half-filled sites in the crystal lattice. Temperature-pressure phase diagram for \SPS\ compound and temperature-composition phase diagram for \SPbPS\ mixed crystals, which include tricritical points and where a temperature of phase transitions decrease to 0 K, together with the data about some softening of low energy optic phonons and rise of dielectric susceptibility at cooling in quantum paraelectric state of \PbPS\, are analyzed by GGA electron and phonon calculations and compared with electronic correlations models. The anharmonic quantum oscillators model is developed for description of phase diagrams and temperature dependence of dielectric susceptibility.
    \end{abstract}


\section*{Introduction}
\begin{figure}[!htbp]
\centering
  \includegraphics[width=9cm]{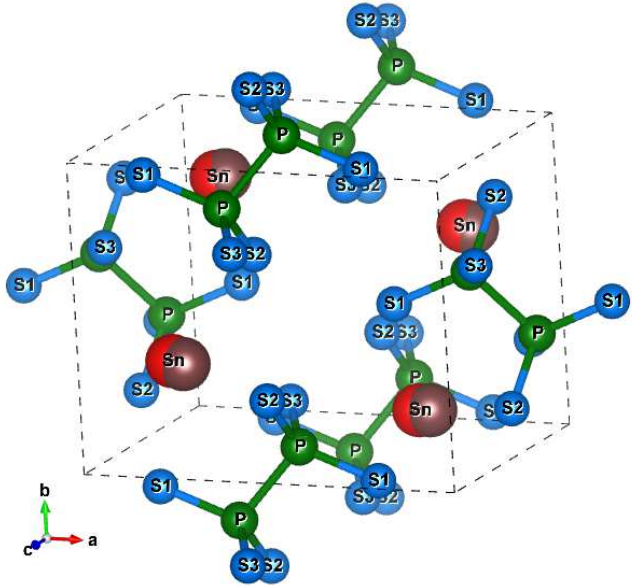}
  \caption{Structure of \SPS\ type crystals in paraelectric (P2$_1$/c) phase with shown shifts of cation atoms in ferroelectric (Pc) phase.}\label{fig1}
\end{figure}

For \SPS\ ferroelectrics, as follows from the first principles calculations~\cite{ref1}, the spontaneous polarization is determined by three-well local potential for the order parameter fluctuations. Continuous phase transition can be described as second order Jahn-Teller (SOJT) effect which is related to the stereoactivity of Sn$^{2+}$ cations placed inside of sulfur ions polyhedron~\cite{ref1,ref2}. For the \SPS\ crystal lattice (Fig.~\ref{fig1}) in mechanism of the ferroelectricity, the hybridization of Sn 5s orbitals and S 3p orbitals is a main driving force.  Thermodynamics of such transition can be described in well-known Blume-Emery-Griffith (BEG) model~\cite{ref3} with two order parameters (dipolar and quadrupolar) and three possible values of pseudospin: -1, 0, +1. The most important future of BEG model is a presence of tricritical point (TCP) on states diagram~\cite{ref3,ref4} -- while the decrease of temperature of phase transition by pressure or by alloying in mixed crystals, the transition becomes of first order. Indeed, below tricritical "waterline temperature" about 220-240~K and at compression above 0.6~GPa, the ferroelectric transition in \SPS\ evolves to first order~\cite{ref5,ref6}. Also, at tin by lead substitution in the \SPbPS\ mixed crystals, for y$>0.3$ and below 220~K, the hysteresis which is related to the paraelectric and ferroelectric phase coexistence appears that give evidence about a discontinuous character of the phase transition~\cite{ref7}.

\begin{figure}[!htb]
\centering
  \includegraphics[width=12cm]{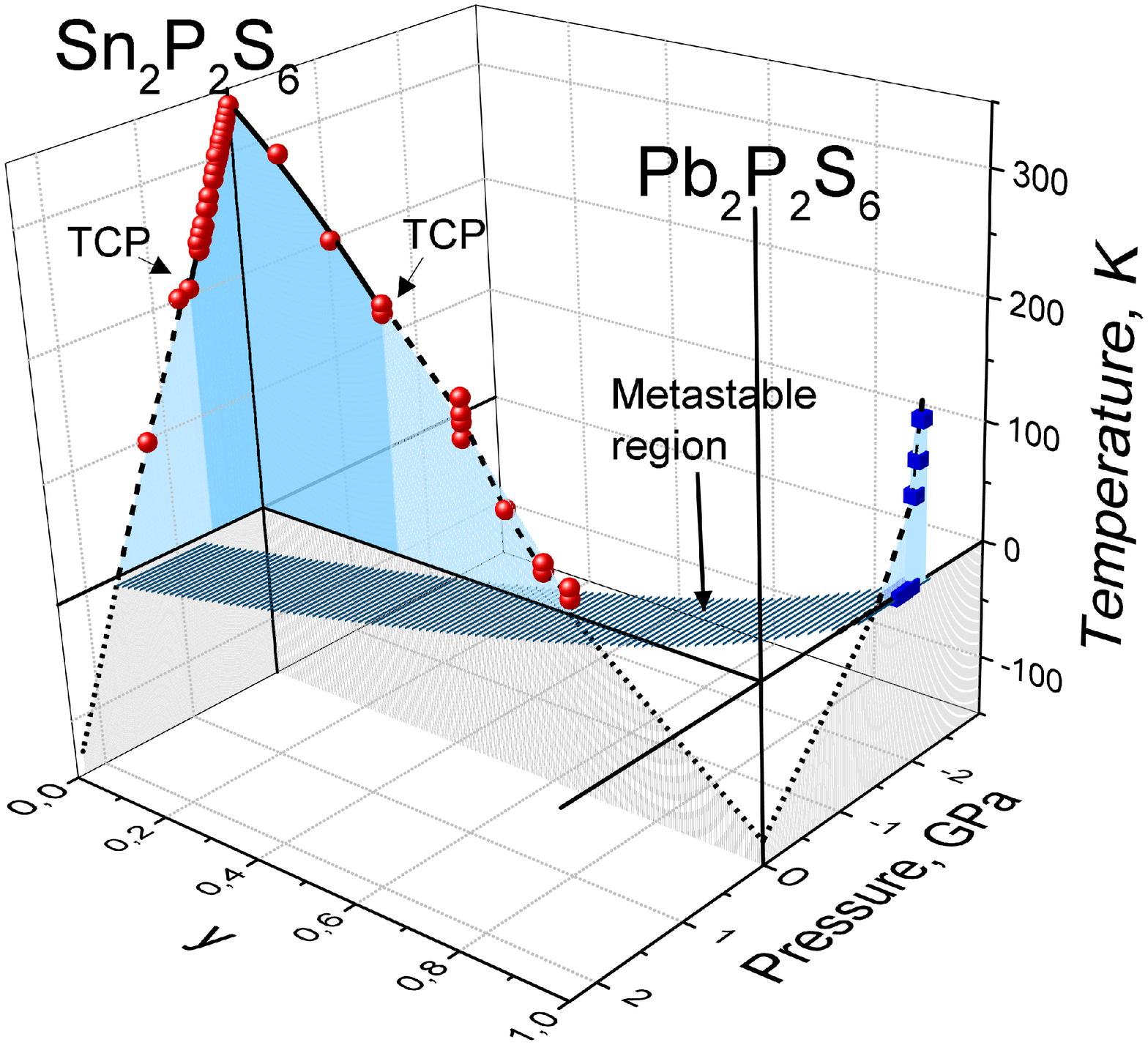}
  \caption{Temperature-pressure phase diagram for \SPS\ crystal and temperature-composition phase diagram for \SPbPS\ mixed crystals. The tricritical points are located near 220~K for pressure $P\approx0.6$~GPa~\cite{ref6} or for concentration $y\approx0.3$~\cite{ref7}. For \PbPS\ crystal the calculated temperature-pressure phase diagram is shown. The metastable region in ground state ($T=0$~K) with possible coexistence of paraelectric and ferroelectric phases is shaded. }\label{fig2}
\end{figure}

One more future of discussed $T-P$ and $T-y$ phase diagrams is finite range of the ferroelectric phase existence. Under compression, the temperature of ferroelectric transition decreases to $T_{\rm C}\sim 110$~K at 1.2~GPa in \SPS\ crystal~\cite{ref6}. The observed $T_{\rm C}(P)$ dependence at linear extrapolation reaches 0~K near the pressure of 1.5~GPa. For the \SPbPS\ mixed crystals the paraelectric phase becomes stable till 0~K above y$\sim$0.7~\cite{ref8}. Comparison of $T_{\rm C}(P)$ and $T_{\rm C}(y)$ dependencies (Fig.~\ref{fig2}) shows, that stability of paraelectric state in \PbPS\ compound can be similar to behavior of \SPS\ crystal under a pressure of 2.2~GPa. With increasing pressure the elastic energy increases what make unfavorable the SOJT effect and ferroelectric state disappears in \SPS\ compound. At transition from \SPS\ to more ionic \PbPS\ compound at normal pressure, the SOJT effect weakens because Pb 6s orbitals have energy about 1~eV lower than in the case of Sn 5s orbitals. This rise of energy distance between S 3p orbitals and Pb 6s orbitals determines suppression of Pb$^{2+}$ cations stereoactivity~\cite{ref2}.

So, according to the experimental data~\cite{ref6,ref8}, the paraelectric ground state can be stable at $P>1.5$~GPa for \SPS, and at $y>0.7$ and normal pressure for \SPbPS\ mixed crystals. At 0~K, the quantum critical point (QCP) can be reached by variation of pressure or composition~\cite{ref9}. Early the QCP have been obviously found in the quantum paraelectric SrTiO$_3$ at increase of $^{18}$O isotope concentration only~\cite{ref10}. Here the ferroelectric phase appears almost continuously at near 45\% content of $^{18}$O, and for SrTi$^{18}$O$_3$ the ferroelectric phase appears below T$_{\rm C}\approx 25$~K. At cooling to QCP the reciprocal dielectric susceptibility is proportional to $T^2$, what is the most important criterion of the quantum critical behavior of susceptibility. But for the most experimental observations, the ferroelectric or magnetic ordering appears by first order transition and $T^2$ behavior of the related susceptibilities is observed only in a some finite temperature interval. Such transformation of QCP into the first order phase transition at $T=0$~K can be explained by an interaction between different order parameters that break different symmetry elements of paraelectric phase~\cite{ref9,ref10}.

For \SPbPS\ mixed crystals with compositions $y\approx0.61$ and $y\approx0.65$, which are closed to the transition at zero temperature from polar phase (y$<$0.7) to paraelectric one (y$>$0.7), the dielectric susceptibility demonstrates the quantum critical behavior $\chi^{-1}\sim T^2$ in vicinity of the first order transitions with $T_{\rm C}\approx35$~K and 20~K, respectively~\cite{ref8}. It is interesting to understand the nature of quantum paraelectric state at high pressures in \SPS\ compound or at high lead concentration in \SPbPS\ mixed crystals.

Appearance of spontaneous polarization at low temperatures can be mostly related to electronic correlations. Early in the extended Falikov-Kimbal (EFK) model it was shown~\cite{ref11,ref12,ref13}, that hybridization between orbitals of localized carriers, at the top of valence band, and itinerant electrons, at the bottom of conduction band, can destroy the crystal lattice symmetry center. In this case the spontaneous polarization in the ground state appears together with excitonic condensate~\cite{ref12}. The EFK model can be related to Hubbard model~\cite{ref14} which also take into account the spins of fermions. It is interesting that the last one is usually converted into anisotropic XYZ Heisenberg model, and for special set of parameters -- into isotropic Heisenberg XXZ model~\cite{ref14}. It should be noted that isotropic Heisenberg model can be equivalent to BEG model~\cite{ref15}, and in fermionic presentation the BEG model is also related to EFK model~\cite{ref11}.

Up to now nobody found experimental evidences about electronic ferroelectricity. A long history of so-called vibronic theory of ferroelectric phase transitions~\cite{ref16} stimulates searching of the electronic origin of crystal lattices spontaneous polarization. The vibronic models are mostly connected to Jahn-Teller effect~\cite{ref17} and they are based on electron-phonon interaction. Obviously the Jahn-Teller effect can be leading at description of compounds with ferroelectric phase transitions at high temperatures. Both, SOJT effect and models of electronic correlations, should be obviously involved for description of ferroelectric systems with three-well local potential. For this case of two order parameters (dipolar and quadrupolar), at lowering of second order transition temperature (by compression or by variation in chemical composition) the TCP is reached. Further the line of first order transitions goes down to 0~K. Here the quantum paraelectric state is evidently realized and possibility of electronic correlation weak effects can be checked. Of course, a some phonon contribution to the electronic correlations is expected, what is accounted in the Holshtein-Hubbard type models~\cite{ref18,ref19}. The relaxations of paired in a real space Anderson's electrons~\cite{ref15,ref20,ref21} or phononic  Kondo screening~\cite{ref22,ref23,ref24} can also be involved in some approaching for description of phonon-like excitations in the low temperature quantum paraelectric state.

In this paper we present the results of dielectric susceptibility and Raman scattering studies of \PbPS\ crystals. Some growth of dielectric susceptibility and softening of the lowest energy phonon optic mode are observed at cooling from room temperature till 20~K in the paraelectric phase of \PbPS. Experimental data on $T-P$ phase diagram of \SPS\ and on $T-y$ phase diagram of \SPbPS\ are compared and involved for explanation of unusual behavior of paraelectric phase of \PbPS\ crystal. The electron and phonon spectra of \PbPS\ crystal are calculated in GGA approximation of Density Functional Theory and compared with experimental data. For this crystal the transition from paraelectric state into ferroelectric one under negative pressure is modeled by first-principles calculations.

Taking into account the ideas about valence skipping~\cite{ref25} and charge disproportionation~\cite{ref26}, the electronic origin of possible spontaneous polarization of \PbPS\ crystal lattice is analyzed in the frame of EFK model~\cite{ref12}. Also the approximations of Hubbard model and related BEG model~\cite{ref15} are used for consideration of simplified picture of half-filled lattice with sites that incorporate the SnPS$_3$ atomic groups with 27 valence electrons. The minimal energetic model is proposed that considers the 26 valence electrons on hybridized bonding orbitals of SnPS$_3$ structure units as core and only one electron is considered at every site. The $\rm{P}^{4+} + \rm{P}^{4+}\rightarrow\rm{P}^{3+} + \rm{P}^{5+}$ charge disproportionation process is energetically favorable for P$^{4+}$ phosphorus ions valence skipping and $3s^1+3s^1\rightarrow3s^2 + 3s^0$ recharging. The on-site Coulomb repulsion $U_{\rm C}$ acts in opposite way to this disproportionation process whith  energy gain $U_{\rm disp}$. The orbitals hybridization complicates this picture, but also induce acentric state. We estimate the energy of disproportionation which is main origin of acentric ground state at condition of suppressed SOJT effect. At  $U_{\rm C}\approx |U_{\rm disp}|\approx E_g\approx 2$~eV, where $E_g$ is a energy gap, the EFK model was used for estimation of electronic contribution to dielectric susceptibility. The calculated electron spectra are compared with parameters of minimal BEG like energetic model that is derived from Hubbard Hamiltonian~\cite{ref15} and agrees with observed $T-P$ and $T-y$ diagrams of \SPbPS\ ferroelectrics. Finally, the quantum anharmonic oscillators models, that is based on statistics of phonon-like excitations in the lattice with three-well local potential, is proposed for description of $T-P$ and $T-y$ state diagrams of \SPbPS\ ferroelectrics and explanation of their dielectric susceptibility growth at cooling in the quantum paraelectric state.

\section{Experimental data}
The dielectric susceptibility and Raman scattering were investigated for \PbPS\ crystals that was grown by vapor transport method. Temperature dependence of dielectric susceptibility was investigated by Bridge scheme at frequencies range $10^4-10^8$~Hz~\cite{ref7}. The Raman scattering was excited by He-Ne laser and analyzed by DFS-24 spectrometer with 1 cm$^{-1}$ frequency resolution. The spectra were fitted by Voight shape spectral counters.

\begin{figure}[!htbp]
\centering
  \includegraphics[width=10cm]{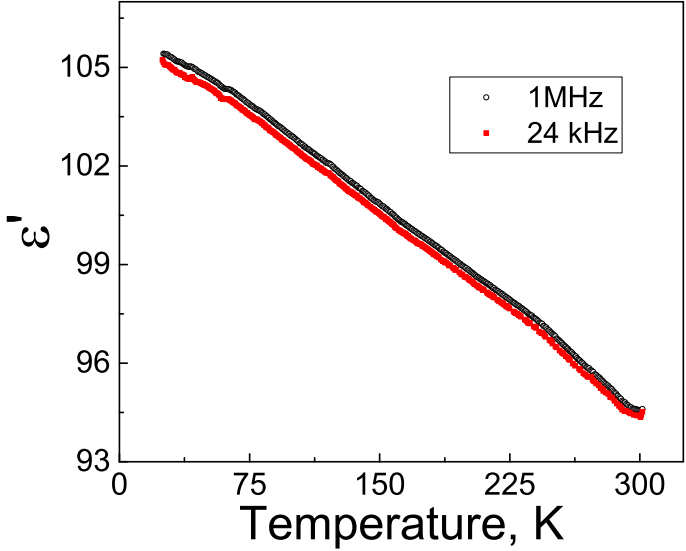}
  \caption{Temperature dependence of \PbPS\ dielectric susceptibility at different frequencies.}\label{fig3}
\end{figure}

The temperature dependence of real part of dielectric susceptibility is shown at Fig.~\ref{fig3} for different frequencies. The susceptibility increases about ten percepts at cooling in the region of three hundred Kelvins without frequency dispersion in the studied diapason.

\begin{figure}[!htb]
\centering
  \includegraphics[width=8cm]{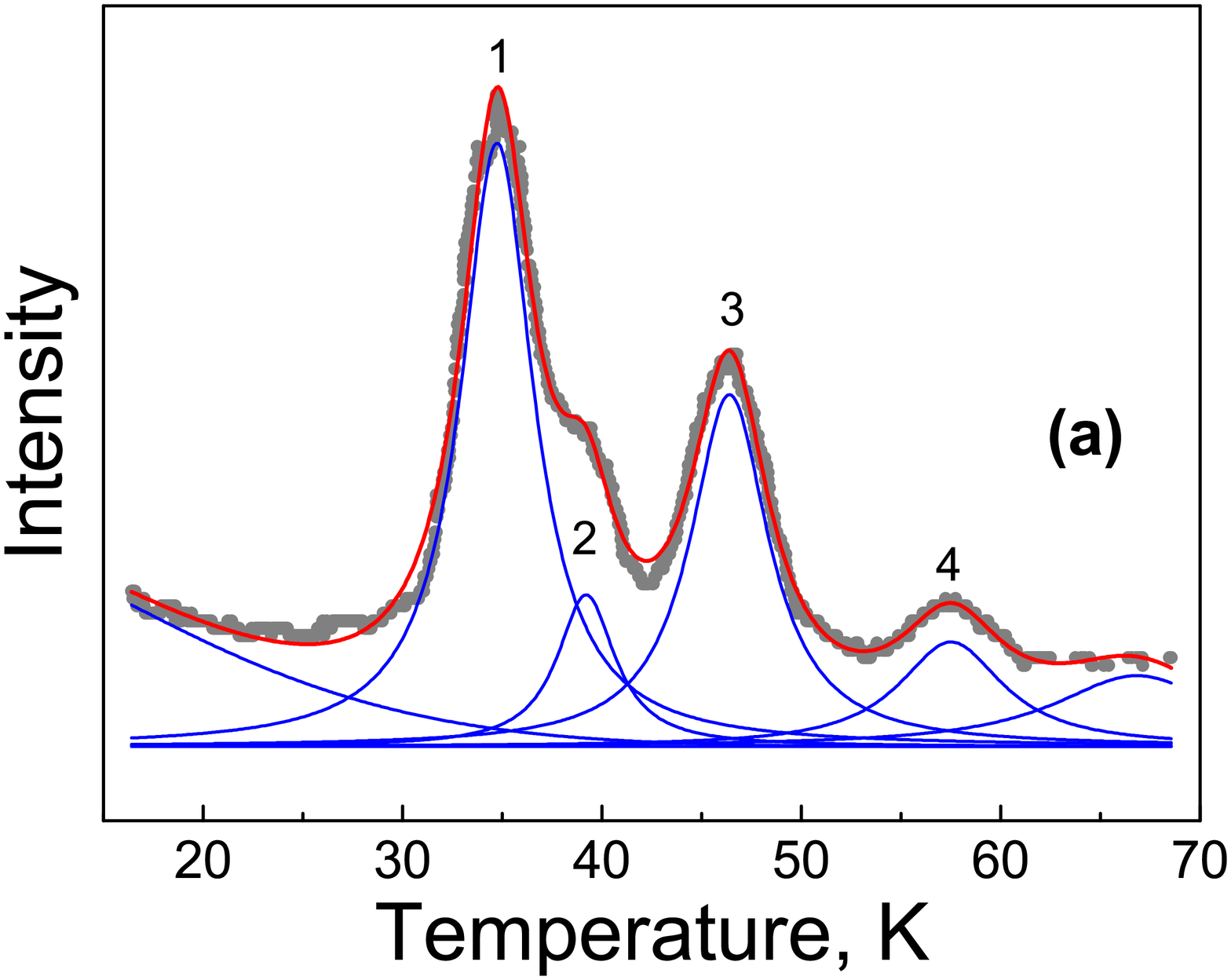}
  \includegraphics[width=8cm]{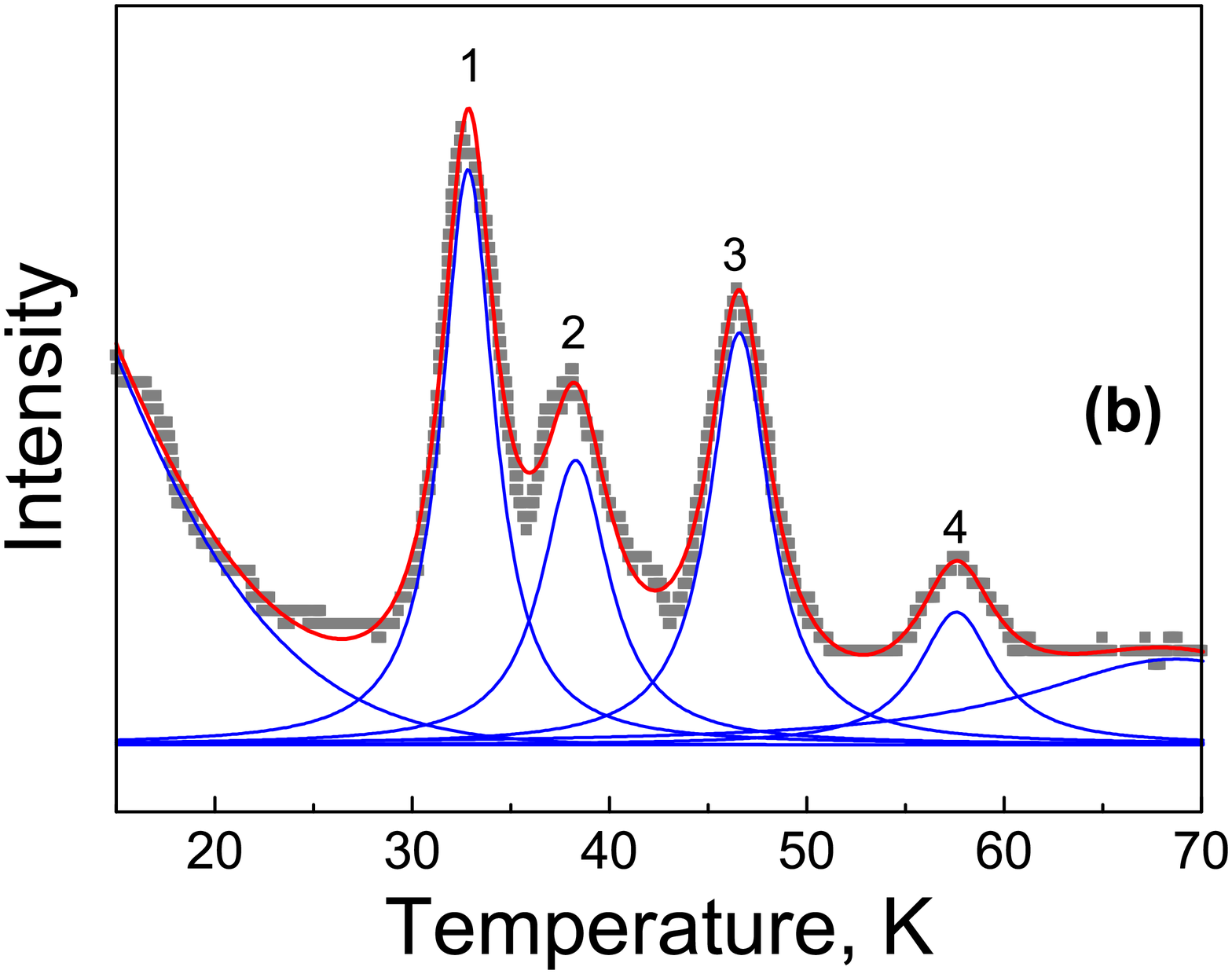}
  \includegraphics[width=8cm]{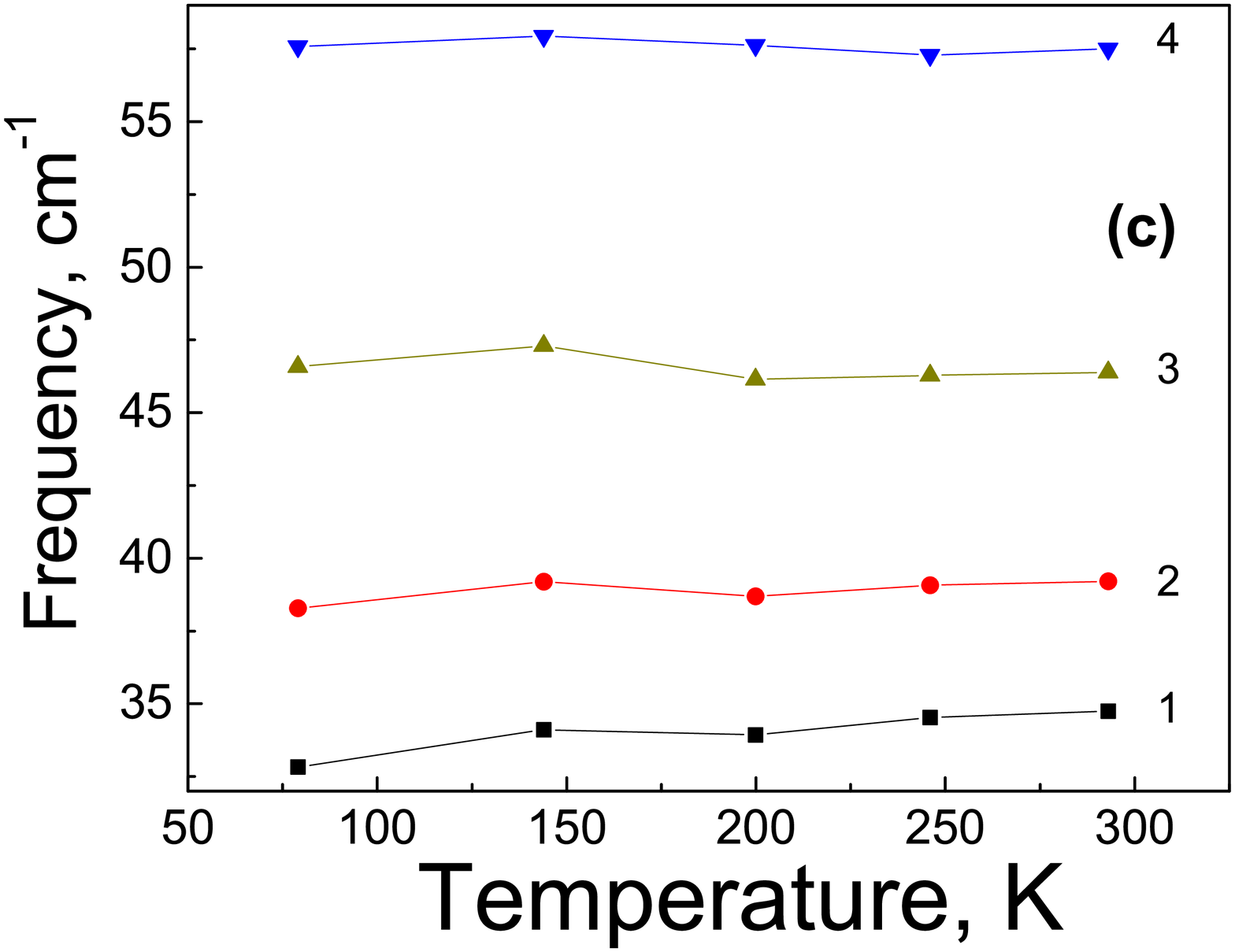}
  \includegraphics[width=8cm]{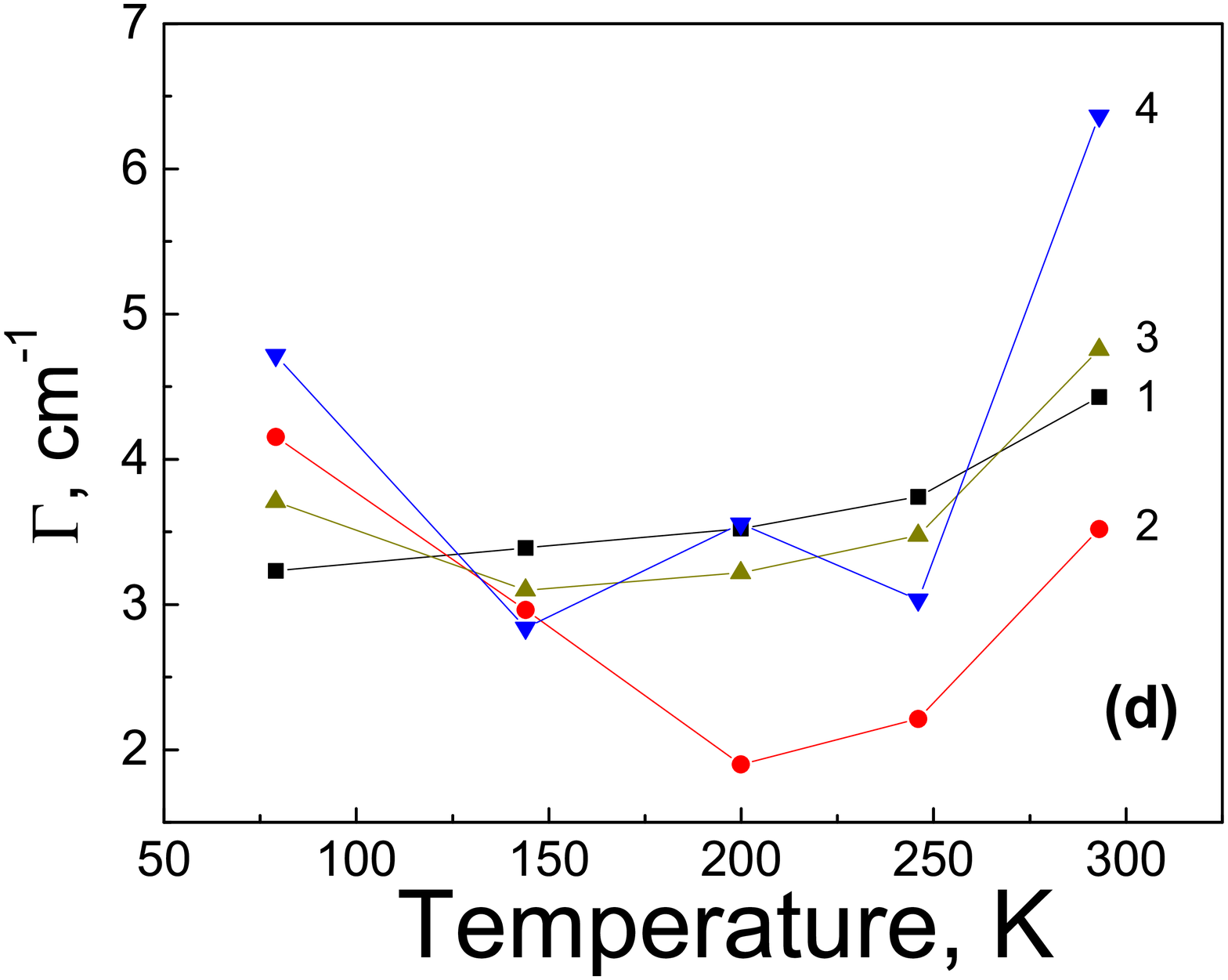}
  \caption{Raman scattering spectra for low energy optic modes in \PbPS\ crystals at 293~K (a) and 79~K (b), and temperature dependence of spectral lines frequencies (c) and damping (d).}\label{fig4}
\end{figure}

According to temperature evolution of Raman spectra (Fig.~\ref{fig4}), the lowest energy optic mode softens about 2~cm$^{-1}$ at cooling from room temperature to 79~K. At temperature decrease, the low frequency spectral tail and damping of optical phonons growth also. Such behavior coincides with earlier found~\cite{ref27} increase of \PbPS\ crystal lattice anharmonicity according to the rise of Gruneisen parameter at cooling on the data of Brillouin scattering by acoustic phonons. The observed rise of dielectric susceptibility about 10\% at cooling correlates with a lowering of optic mode frequency by 1~cm$^{-1}$.

\section{Calculation of energetic spectra}
\begin{figure*}[!htb]
\centering
  \includegraphics[width=14cm]{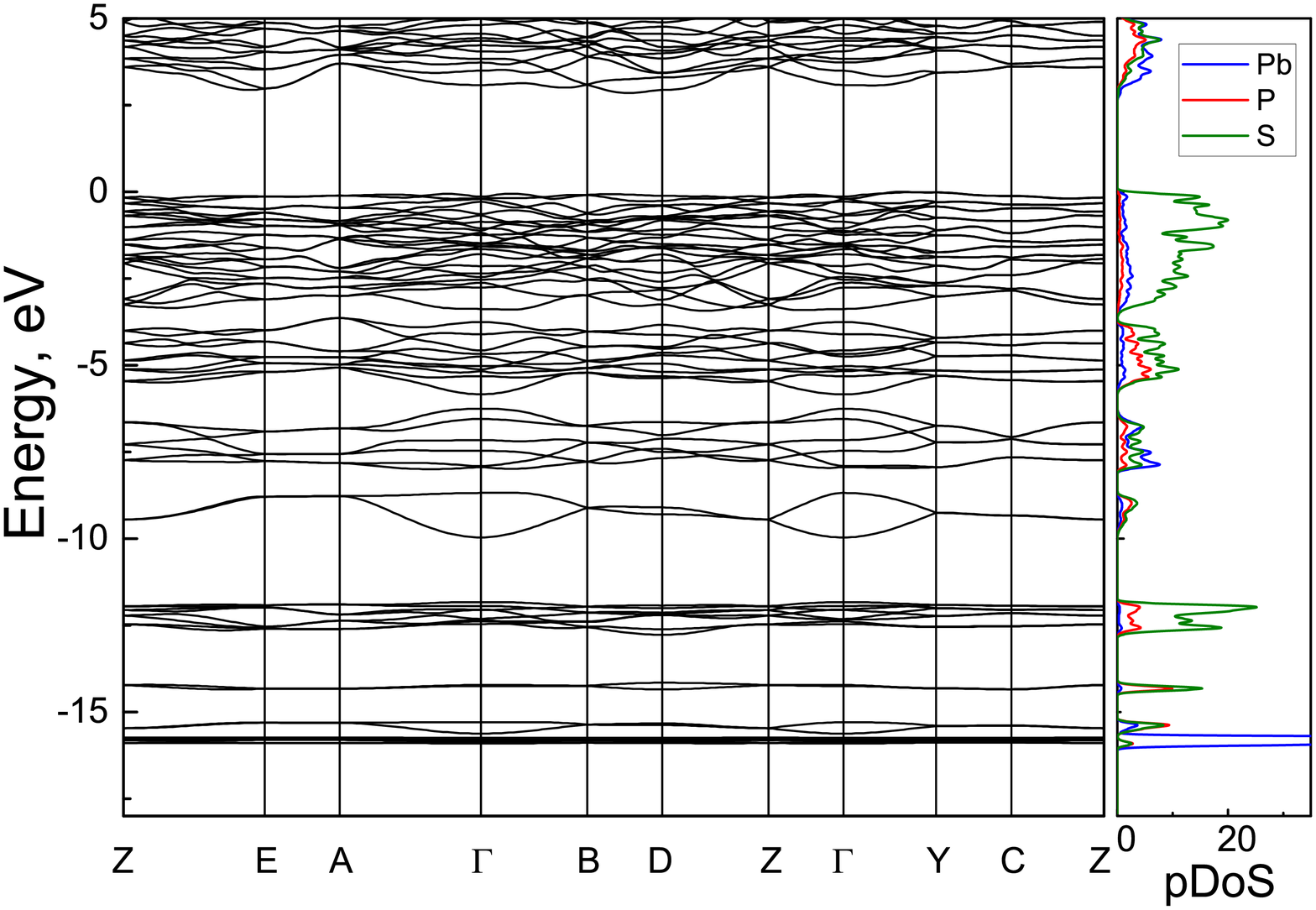}
  \caption{The electron band structure and partial electron densities of states for \PbPS\ crystal.}\label{fig5}
\end{figure*}

For characterization of \PbPS\ paraelectric phase and to search a possible origin of its instability, the electron and phonon spectra were calculated in GGA approach of DFT. The monoclinic elementary cell (Fig.~\ref{fig1}) contains two formula units and belong to P2$_1$/c space group~\cite{ref28,ref29}. Structure of valence zone (Fig.~\ref{fig5}) correlates with earlier~\cite{ref2} described for \SPS\ analog. But here the contribution of Pb$^{2+}$ cations 6s orbitals into the density of states at the top of valence band is smaller as in the case of Sn$^{2+}$ 5s orbitals presence. This fact reflects smaller stereoactivity of Pb$^{2+}$ cations in more ionic \PbPS\ lattice. Such peculiarity is evidently determined by bigger energy distance between sulfur 3p orbitals and lead 6s orbitals. Indeed, the Pb 6s orbitals are placed mostly below -8~eV -- by 1~eV lower in compare with energetic position of Sn 5s orbitals in \SPS\ crystal.

It is important to mention that for both edges of valence band and conductivity band, the orbitals of sulfur, phosphorous and lead atoms are presented. Obviously the electronic structure of \PbPS\ compound can be considered as bounding and antibounding counterparts of PbPS$_3$ atomic groups orbitals.

The electron spectra calculations in GGA approach are in good enough agreement for the energy gap value - calculated one (about 2.2~eV) is closed to 2.45~eV gap according to optic absorption measurements at 4.2~K~\cite{ref30}.

\begin{figure}[!htbp]
\centering
  \includegraphics[width=10cm]{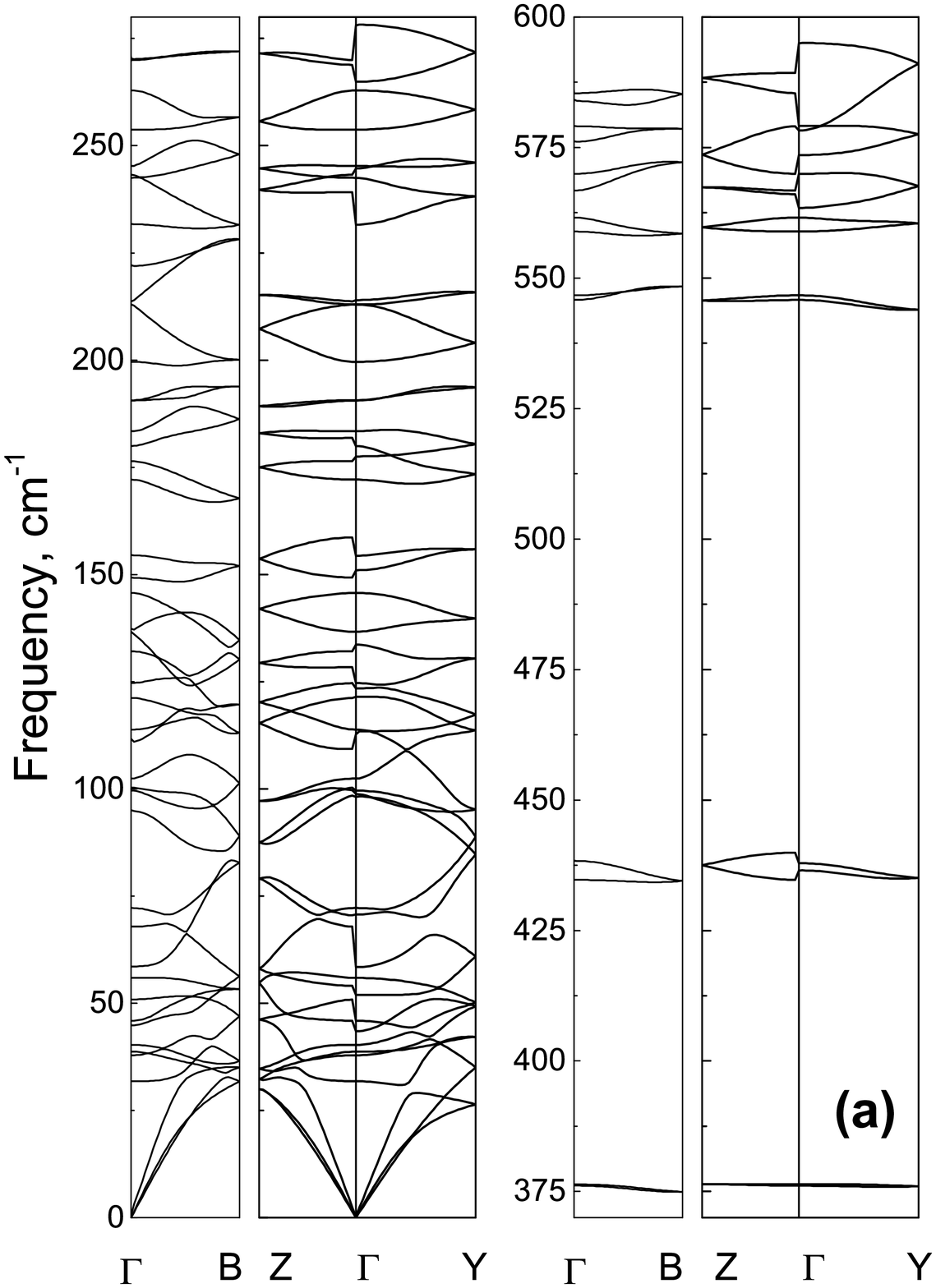}\\
  \includegraphics[width=10cm]{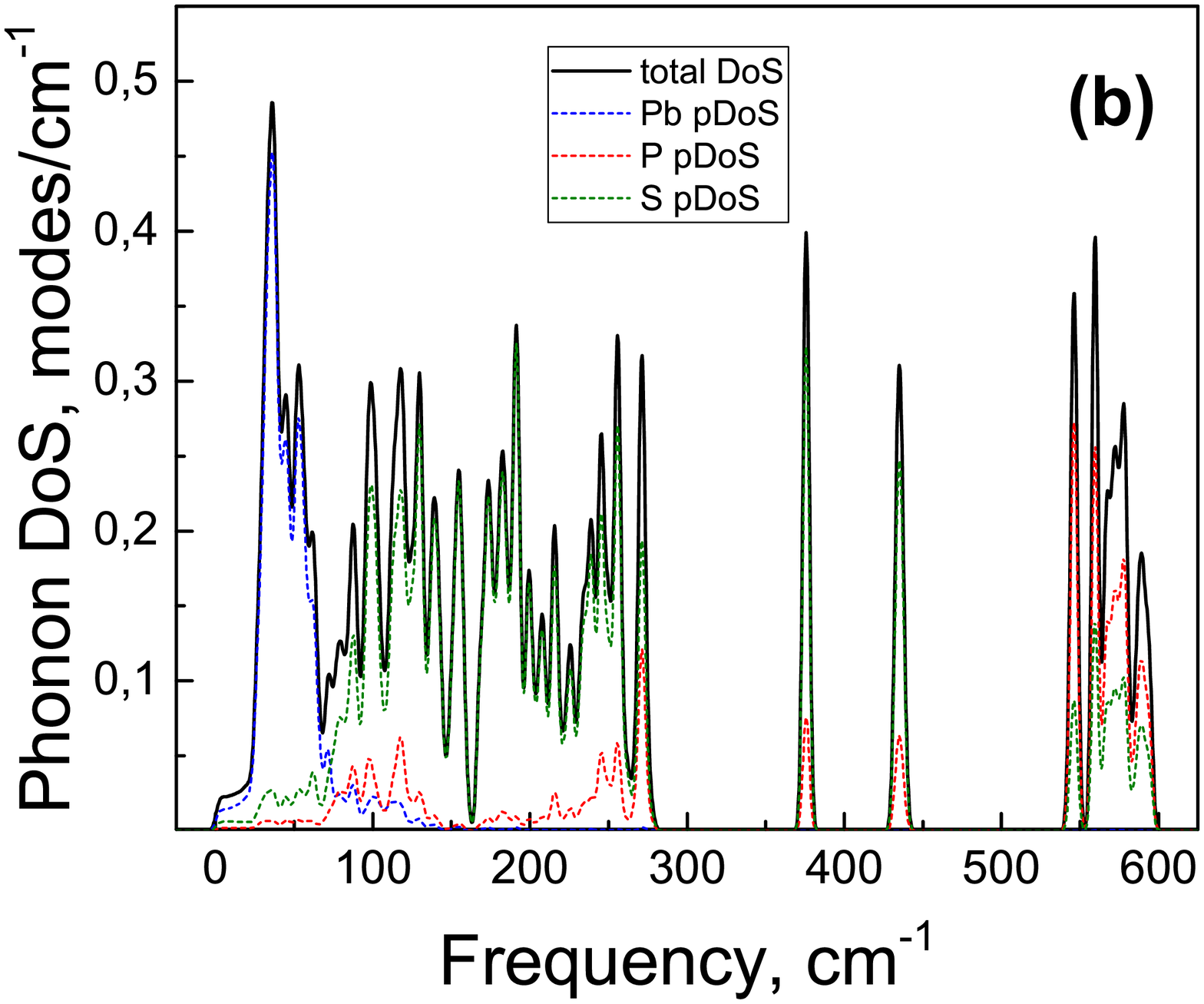}
  \caption{Calculated phonon spectra (a) and partial phonon densities of states (b) for \PbPS\ crystal. }\label{fig6}
\end{figure}

In GGA approach the phonon spectra for \PbPS\ crystal lattice were calculated also. The phonon branches and partial phonon densities of states are presented at Fig.~\ref{fig6}. Long wave phonon frequencies are in good agreement with experimental data what give evidence about adequate GGA approach for the energetic spectra analysis. Below 70~cm$^{-1}$, the translations of Pb ions relatively to P$_2$S$_6$ structural groups mostly contribute into phonon eigenvectors as follows from the partial phonon densities of states. Above 150~cm$^{-1}$ the internal vibrations of (P$_2$S$_6$)$^{4-}$ anions dominate and in higher frequency interval 530-590~cm$^{-1}$ mostly phosphorous atoms participate in the phonon eigenvectors.

The LO-TO splitting has the biggest values for the translational lattice polar modes with frequencies 43, 58 and 70~cm$^{-1}$. It is remarkable that big LO-TO splitting also was found for the (P$_2$S$_6$)$^{4-}$  internal valence vibrations with frequencies 563, 566 and 578~cm$^{-1}$.

\begin{figure}[!htbp]
\centering
  \includegraphics[width=10cm]{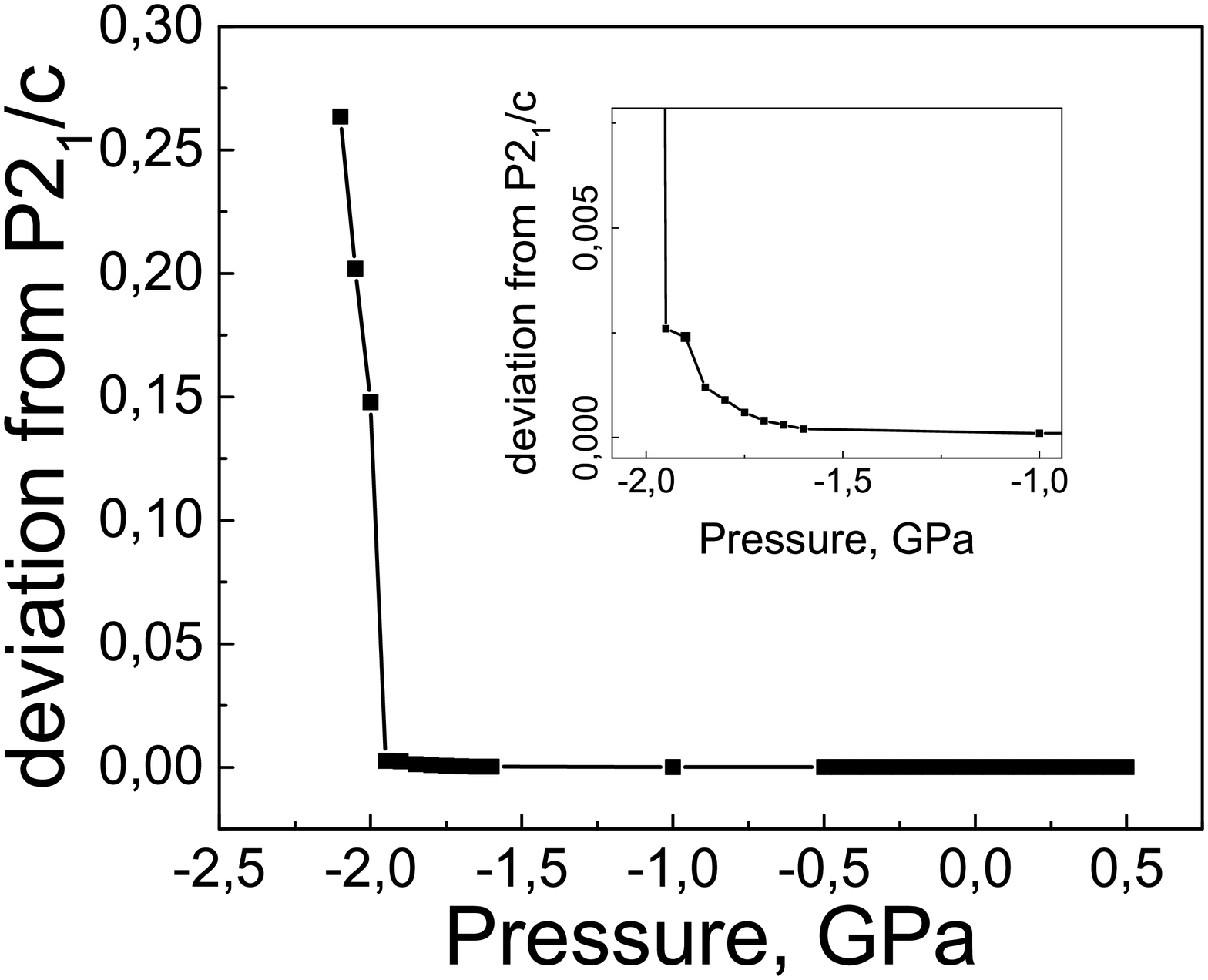}
  \caption{Calculated acentric deviation induced by negative compression of \PbPS\ crystal lattice.}\label{fig7}
\end{figure}

Adequate calculations of electronic and phonon spectra of \PbPS\ crystal at normal pressure permit to suppose a possibility for obtaining of reasonable calculation results at negative pressure. It was found that the transition into ferroelectric phase occurs near -1.95~GPa (Fig.~\ref{fig7}).  It is important that the acentric deviation appears already at pressure about -1.7~GPa.  The ground state of \PbPS\ crystal at normal pressure is similar to the \SPS\ ground state at 2.2~GPa (Fig.~\ref{fig2}).

Some similarity of \SPS\ compound $T-P$ diagram at pressures $0-2.2$~GPa with calculated states diagram for \PbPS\ crystal in pressure range from -2 to 0~GPa permits us to suppose that temperature behavior of \PbPS\ crystal (at normal pressure) could be similar to \SPS\ temperature behavior at 2.2~GPa. This behavior can be determined by some polar fluctuations, like in quadrupole phase according to MC simulations \cite{ref31}, and some growth of dielectric susceptibility at cooling can be expected. Let's try to found some explanation for quantum paraelectric behavior in \PbPS\ crystal.

\section{Discussion of spontaneous polarization origin}
Earlier~\cite{ref1,ref2}, for \SPS\ ferroelectrics the stereoactivity of Sn$^{2+}$ cations, which are placed inside sulfur octahedron, was only considered as origin of local dipoles that exist already in paraelectric phase, because tin cations occupy general position in the monoclinic elementary cell. This stereoactivity is determined by covalency between tin atomic orbitals and molecular orbitals of P$_2$S$_6$ groups, and can be considered as second order Jahn-Teller effect, which involve phosphorous orbitals also. At weaking of SOJT effect by pressure, or at tin by lead substitution, another origins of crystal acentricity can be observed. The first of them is disproportionation of P$^{4+}$ cations: $\rm{P}^{4+} + \rm{P}^{4+}\rightarrow\rm{P}^{3+} + \rm{P}^{5+}$. This process is also related to recharging of surrounding sulfur anions that is accounted by polarizability of nearest space~\cite{ref26}.

The energy of disproportionation can be determined as $U_{\rm disp}=U_{\rm vac}-U_{\rm pol}$. Here $U_{\rm vac}=I_5-I_4$ is the difference in ionization potentials for P$^{5+}$ and P$^{4+}$ ions in vacuum. The polarization energy is determined by Born formula~\cite{ref26} $U_{\rm pol}=(e^2/2R^+)(1 - \gamma/\epsilon_{\infty})$, where $\gamma=1-(R^+/(R^++R^-))(1- \epsilon_{\infty}/\epsilon_0)$. Here the phenomenological parameters are included -- anions and cations radii $R^-$ and $R^+$, respectively, and dielectric constants $\epsilon_0$ and $\epsilon_{\infty}$.

The value $U_{\rm vac}\approx13.1$~eV was estimated by using the ionization energies $I_5=62.74$~eV for P$^{5+}$ cations and  $I_4=49.63$~eV for P$^{4+}$ cations~\cite{ref32}. The value $\gamma\approx0.82$ was found for next set of parameters: S$^{2-}$ anion radius $R^-\approx2.03$~{\AA}, according S-P interatomic distance~\cite{ref28}, and P$^{4+}$ cation radius $R^+\approx0.42$~{\AA}; dielectric constants  $\epsilon_0\approx50$ and $\epsilon_{\infty}\approx7.84$~\cite{ref33}. At such parameters, the polarization energy $U_{\rm pol}\approx-15.1$~eV have been found and, sequently, $U_{\rm disp}\approx-2$~eV.

Negative value of $U_{\rm dispr}$ reflects the valence skipping effect that favors 3s$^0$ and 3s$^2$ electronic configuration of P$^{5+}$ and P$^{3+}$ cations instead of 3s$^1$ configuration of P$^{4+}$ ions.

The phosphorous ions charge disproportionation, or valence ordering process, can be viewed as a lattice of Anderson electron pairs~\cite{ref20} which is stabilized by polarizing of surrounding sulfur polyhedrons. The energy to displace each S atom by $x$ can be written as~\cite{ref15}
$E_{ab}=Cx^2/2-gx(q_a-q_b)$ where $q_a-q_b$ is the charge difference between the two nearest neighbor phosphorous atoms, $g$ and $C$ are coupling constants. The Hubbard type Hamiltonian~\cite{ref15} can be applied for description of electrons hopping in band with contribution of phosphorous orbitals. Such Hamiltonian contains the intra-site and inter-site Coulomb interactions. The first one is determined by $U_{\rm C}$ constant. The inter-site interaction for simplicity can be presented by the short-range interaction $e^2\alpha/6\epsilon a$ only, where $\alpha$ is Madelung constant, $a$ is lattice parameter, $e$ is charge of electron and $\epsilon$ is dielectric permittivity.

At condition of small band width limit in the spin presentation the Hubbard model can be reflected onto the BEG model~\cite{ref4,ref15}:
\begin{equation}\label{eq1}
    H=\Delta\sum_i m_i^2+J\sum_{<ij>}m_im_j.
\end{equation}

\begin{figure}[!htbp]
\centering
  \includegraphics[width=10cm]{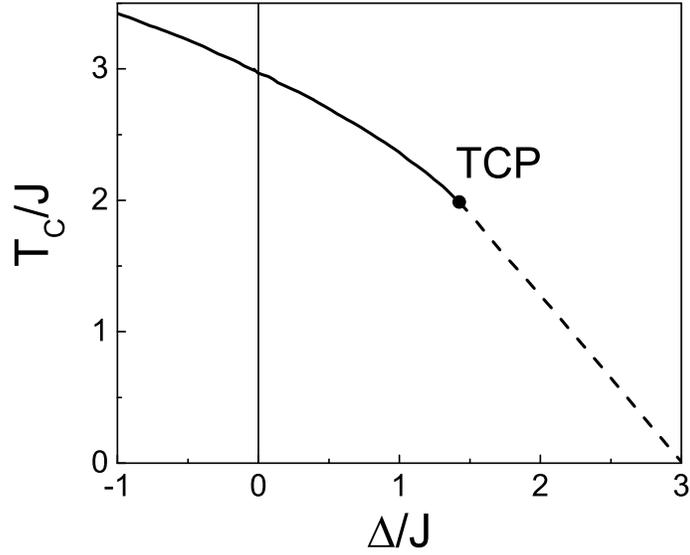}
  \caption{The phase transition temperature as a function of $\Delta/J$ calculated in the mean-field approximation on the BEG model (1). Solid lines denote second order and dashed lines first order transitions and the dot, a tricritical point~\cite{ref15}.}\label{fig8}
\end{figure}

Here parameters are derived from Hubbard model of Anderson's electron pairs -- $\Delta=\frac12(U_{\rm C}-6g^2/C)$ and $J=g^2/C+e^2\alpha/6\epsilon a$. Pseudospin variable $m_i$ has values +1, 0, -1 and can be are related to P$^{3+}$, P$^{4+}$ and P$^{5+}$ states of phosphorous cations. General view of calculated for BEG model phase diagram (Fig.~\ref{fig8}) correlates with experimental observations. Indeed, under compression the lattice stiffness constant $C$ increases, lattice period $a$ lowers and following the on-site energy $\Delta$ rises and the intersite interaction energy $J$ remains almost unchanged.

At tin by lead substitution, the stiffness decrease a little~\cite{ref27}, but the lattice period demonstrates some rise~\cite{ref28}. So again, the intercell interaction $J$ remains almost unchanged. But in lead compound the ionicity is bigger than in the case of \SPS, and the onsite Coulomb repulsion $U_{\rm C}$ is obviously higher. At this the on-site energy $\Delta$ also growth at Sn by Pb substitution. Such influence of chemical composition on the system ground state qualitatively agrees with derived in Hubbard approximation~\cite{ref34} phase diagram where phase transition from paraelectric phase into state with dipole ordering is governed by difference between energies of filled orbitals of two type cations in the crystal lattice. In our case, according to GGA calculations (Fig.~\ref{fig5}), the difference between P 3p orbitals energy and Sn 5s or Pb 6s orbitals energies increases nearly 1~eV at transition from \SPS\ to \PbPS. Evidently the growth of ionicity stabilizes the paraelectric phase what agrees with experimental observation.

Finally, we can estimate some parameters by using the following characteristics: the second order phase transition temperature ($T_0\approx337$~K) for \SPS\ crystal, the coordinates of tricritical points ($T_{\rm TCP}\approx220$~K) on $T-y$ and $T-P$ diagrams, the composition $y\approx0.7$ or pressure $P\approx1.5$~GPa at which the phase transition temperature goes down to zero. By comparing the experimental diagram (Fig.~\ref{fig1}) with calculated one (Fig.~\ref{fig8}), it can be found that $J\approx110$~K for considered ferroelectrics. At almost constant value of intersite interaction energy (parameter $J$), the on-site energy parameter $\Delta$ mostly determines the global temperature-pressure-composition phase diagram. The $\Delta$ parameter has small value and is governed by balance of two significant characteristics -- the on-site Coulomb repulsion $U_{\rm C}$ and disproportionation energy $U_{\rm disp}$. The first one have the value above 2~eV according to {\it ab-initio} GGA+U calculations~\cite{ref35} that explain observed optic gap width $E_g\approx2.4$~eV for \SPS\ crystal. The disproportionation energy $U_{\rm disp}$ can be compared with ratio $6g^2/C$ and equals about -2~eV. Obviously, for \SPS\ crystal a some small value of $\Delta$ can be found (Fig.~\ref{fig8}) and $\Delta\approx0$ for some pressure $P$ or lead content $y$. At $\Delta\approx0$ or for condition $U_{\rm C}\approx U_{\rm disp}$, the valence fluctuations are strongly developed, and possibly, they are still enough strong at rise of $\Delta$ value to $3J$ and phase transition temperature $T_{\rm C}$ going down to 0~K. So, the valence or charge fluctuations can contribute to growth of dielectric susceptibility for \SPbPS\ crystals with $y>0.7$ and for \PbPS\ compound.

Recent calculations in Hubbard model show~\cite{ref34} that in paraelectric ground state, the charge density at different cations (phosphorous and lead in our case) strongly depends on temperature. This peculiarity can be evidently related to the temperature dependence of dielectric susceptibility.

Fluctuations of Anderson's 2e electron pairs as origin of dielectric susceptibility growth at cooling in wide temperature interval can be compared with Kondo screening phenomena~\cite{ref21,ref22,ref23,ref24}. At incoherent 2e pairs fluctuations the susceptibility increases with cooling as $\sim T^{-1}$, but below some temperature the coherence of electron pairs fluctuations can appear and susceptibility reaches some constant value. The model of Kondo effect with involving of phonon channel definitely predicts~\cite{ref24} a rise of dielectric susceptibility at the system cooling.

At low enough temperatures, near 0~K, electronic correlations themselves can induce the system acentricity. The electronic ferroelectricity has been predicted~\cite{ref12,ref13} as result of hybridization of itinerant electron wave function with hole at the top of valence band and appearance of excitonic condensate. The excitonic order parameter correlates with acentric space distribution of the charge density, as it is follows from the EFK model. For ground state in paraelectric phase at approaching to transition into polar phase, by variation of itinerant electrons energy level $E_f$, the dielectric susceptibility is expected to growth according relation~\cite{ref12}:
\begin{equation}\label{eq2}
    \epsilon=\frac{2N\mu_z^2}{\Omega}\frac{{\rm arccoth}(\frac{E_f}{W})}{W-U_{\rm C}\cdot {\rm arccoth}(\frac{E_f}{W})} ,
\end{equation}
\noindent where $\mu_z$ is the interband dipole element, $\Omega$ is the volume of unit cell, $N$ is the number of sites, $2W$ is the bandwidth. It was taken $W=0.02$~eV and $E_f=2$~eV according to estimations which were made from GGA calculations (Fig.~\ref{fig5}).

As was mentioned above, for \SPS\ crystals at normal pressure very delicate energetic balance $U_{\rm C}\approx U_{\rm disp}$ serves for development of charge fluctuations. Moreover, the energy gap between valence and conductivity gap $E_{\rm g}\approx2.45$~eV is also comparable with estimates values $\approx2$~eV for $U_{\rm C}$ and $|U_{\rm disp}|$. Such conditions permit expect possibility of electronic origin for increased at cooling dielectric susceptibility in paraelectric phase.  Estimation of dielectric susceptibility for \PbPS\ at $T=0$~K within EFK model (2) gives value $\epsilon\approx100$ for the interband transition dipole moment $\mu\approx3e\AA$. At heating the susceptibility will decrease because of thermal disordering of the local electric dipoles.

\section{Anharmonic quantum oscillators model}
Now we propose to use the AQO model~\cite{ref36} that consider phonon-like bosonic excitations. These excitations can be regarded as incorporation of electron-phonon interaction in generalized Holshtein-Hubbard model~\cite{ref18} or in Bose-Habbard model~\cite{ref37}.

In the AQO model, crystal is representing as one dimensional system of anharmonic oscillators which interact via quadratic interaction term. So, Hamiltonian of such a system is
\begin{equation}\label{eq1}
    H=\sum_i\left(T(x_i)+V(x_i)\right)+\sum_{ij}J_{ij}x_ix_j ,
\end{equation}
\noindent where $T(x_i)$ and $V(x_i)$ are operators of kinetic and potential energy, respectively, $x_i$ is a $i$ oscillator's displacement, $J_{ij}$ are coupling constants between $i$ and $j$ oscillators. In the mean-field approach, the last term of (\ref{eq1}) can be replaced by another one - $\sum_iJ\langle x\rangle x_i$ where $\langle x\rangle$ is average position of all other oscillators in the lattice. In this case, the Hamiltonian~(\ref{eq1}) can be represented as a sum of independent single-particle Hamiltonians:
\begin{eqnarray}\label{eq2}
\nonumber   H&=&\sum_iH^{eff}_i, \\
  H^{eff}_i&=&T(x_i)+V(x_i)+J\langle x\rangle x_i .
\end{eqnarray}
\noindent Such effective particle is an oscillator under the influence of linear symmetry-breaking field which is calculated self-consistently. Solving Schr\"{o}dinger equation with Hamiltonian (\ref{eq2}) one obtain a set of eigen energies $\{E_n\}$ of levels and its wave functions $\{\Psi_n(x)\}$ which are used for self-consistent calculation of average expectation value $\langle x\rangle$. In our calculations we used matrix representation form for position and momentum operators to solve Schr\"{o}dinger equation~\cite{korch}.
\begin{figure}[!htbp]
\centering
  \includegraphics[width=9cm]{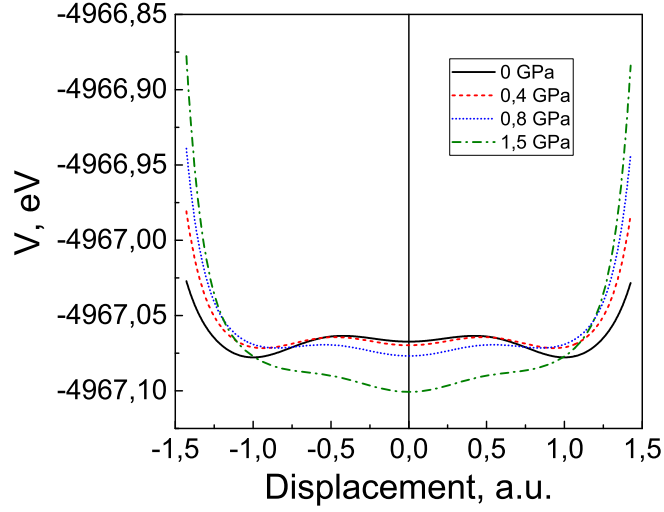}
  \caption{The shape of local potential for \SPS crystal at different pressures~\cite{ref31}. }\label{fig10}
\end{figure}

It should be noted that temperature enters in Hamiltonian (\ref{eq2}) indirectly through Boltzmann distribution for the occupation of the levels. Really, from definition of $\langle x\rangle$ it follows
\begin{equation}\label{eq3}
    \langle x\rangle=\sum_n p_n x_n ,
\end{equation}
\noindent where $p_n\sim exp(-E_n/kT)$ is the occupation number for $n^{th}$ level, $k$ is a Boltzmann constant, $x_n=\int \Psi_n^*x\Psi_ndx/\int \Psi_n^*\Psi_ndx$ is average value of displacement for $n^{th}$ level.

Obviously, in the paraelectirc phase for average desplacement, which is proportional to a order parameter, one obtain $\langle x\rangle=0$. On the other hand, below phase transition temperature $T_c$ we have $\langle x\rangle\neq0$. So, one can model a changes on phase diagram at ionic substitution or pressure by varying coupling constant or shape of potential energy.  Here we use a transformation of the local three-well potential $V(x_i)$ (Fig.~\ref{fig10}) as function of pressure according to the results of~\cite{ref31}. In present calculations the mass of oscillator was equal to tin atomic mass.  Moreover, it is possible to calculate a dielectric response of such the system of oscillators using following relation~\cite{vaks}:
\begin{equation}
\epsilon(\omega,T)=1+\frac{e^2_{eff}}{\epsilon_0V}\left(\frac{\Pi(\omega,T)}{1-J\Pi(\omega,T)}\right),
\end{equation}
\noindent where $ \Pi(\omega,T)=\sum_{\alpha\beta}\frac{(p_{\alpha}-p_{\beta})|\langle\Psi_{\alpha}|x|\Psi_{\beta}\rangle|^2}{\omega-\omega_{\alpha}+\omega_{\beta}-i\omega\gamma}$, and effective charge of oscillator $e_{eff}$ and its volume $V$ can be used as fitting parameters.

\begin{figure}[!htbp]
\centering
  \includegraphics[width=8.cm]{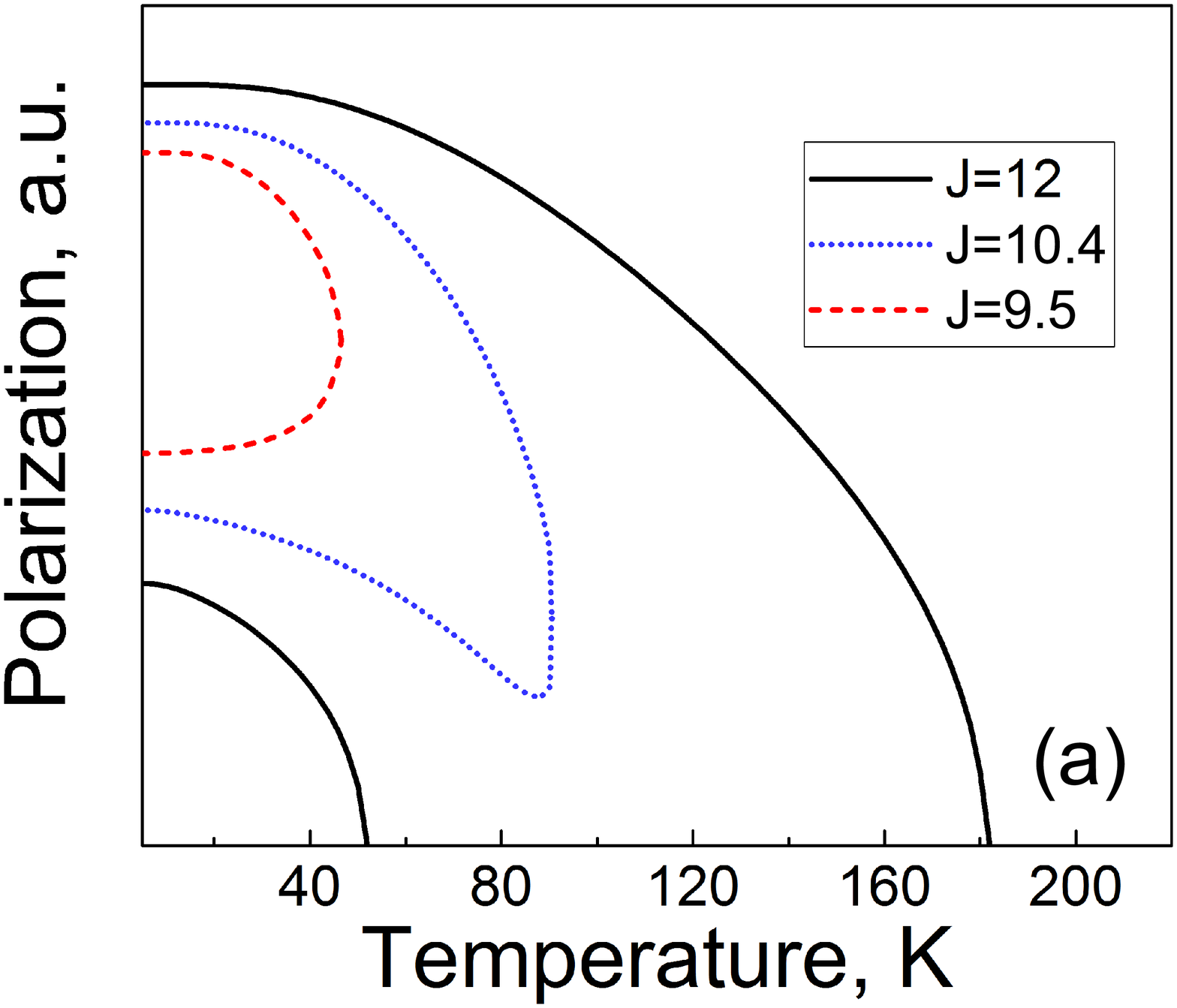}%
  \includegraphics[width=8.cm]{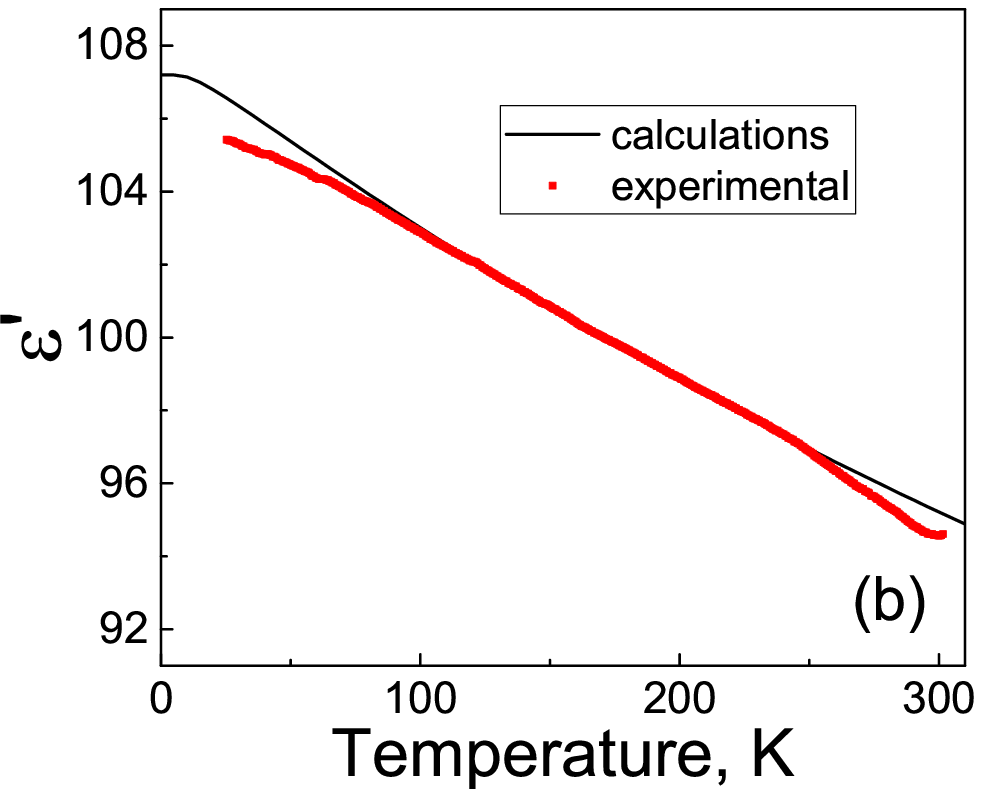}
  \caption{The examples of spontaneous polarization temperature dependence for AQO system for different values of coupling constant $J$ at pressure $p=1.5$~GPa (a); calculated (continuous line) and experimental (points) temperature dependence of dielectric susceptibility for \PbPS\ quantum paraelectric (b).}\label{fig11}
\end{figure}

\begin{figure}[!htbp]
\centering
  \includegraphics[width=10cm]{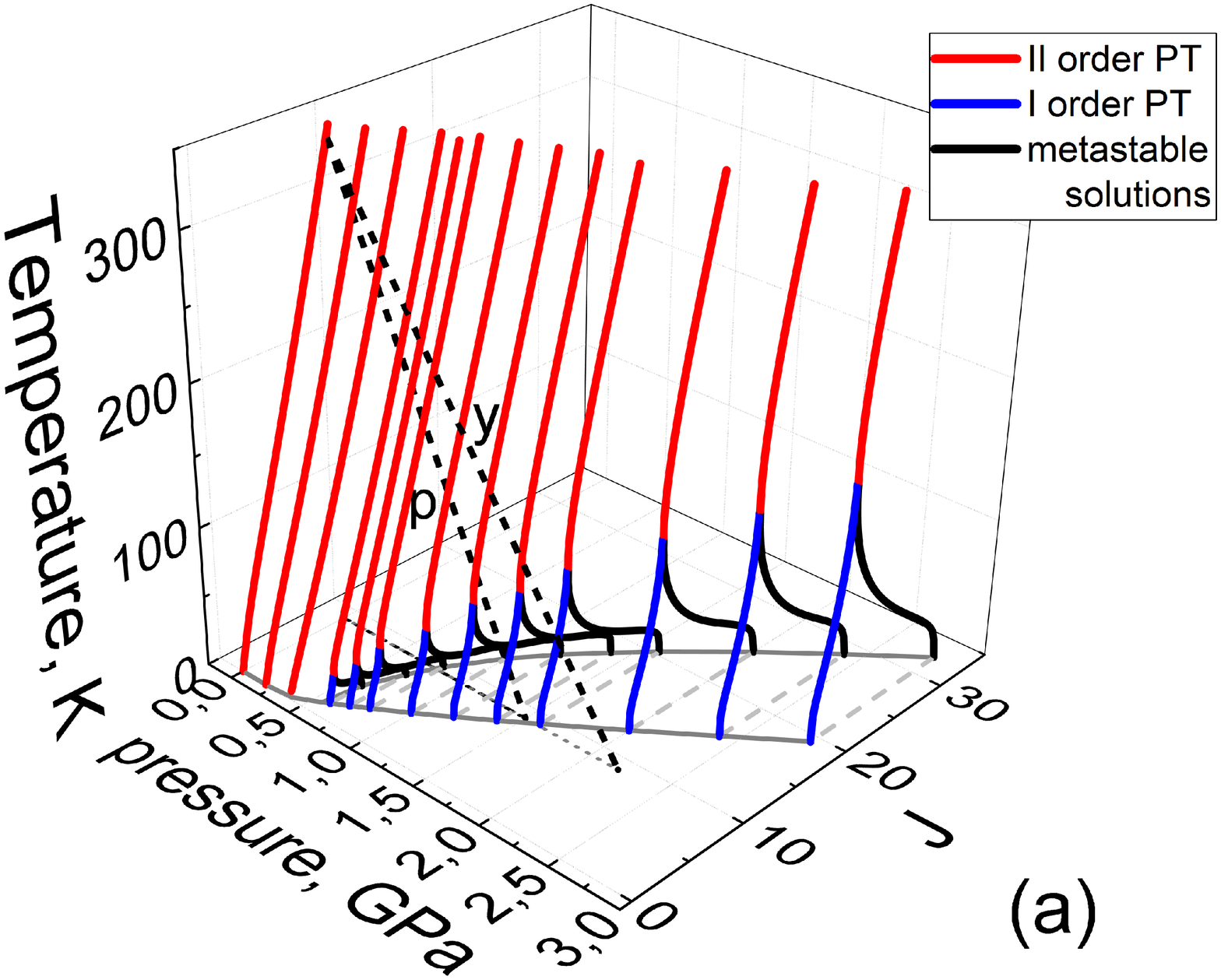}
  \includegraphics[width=10cm]{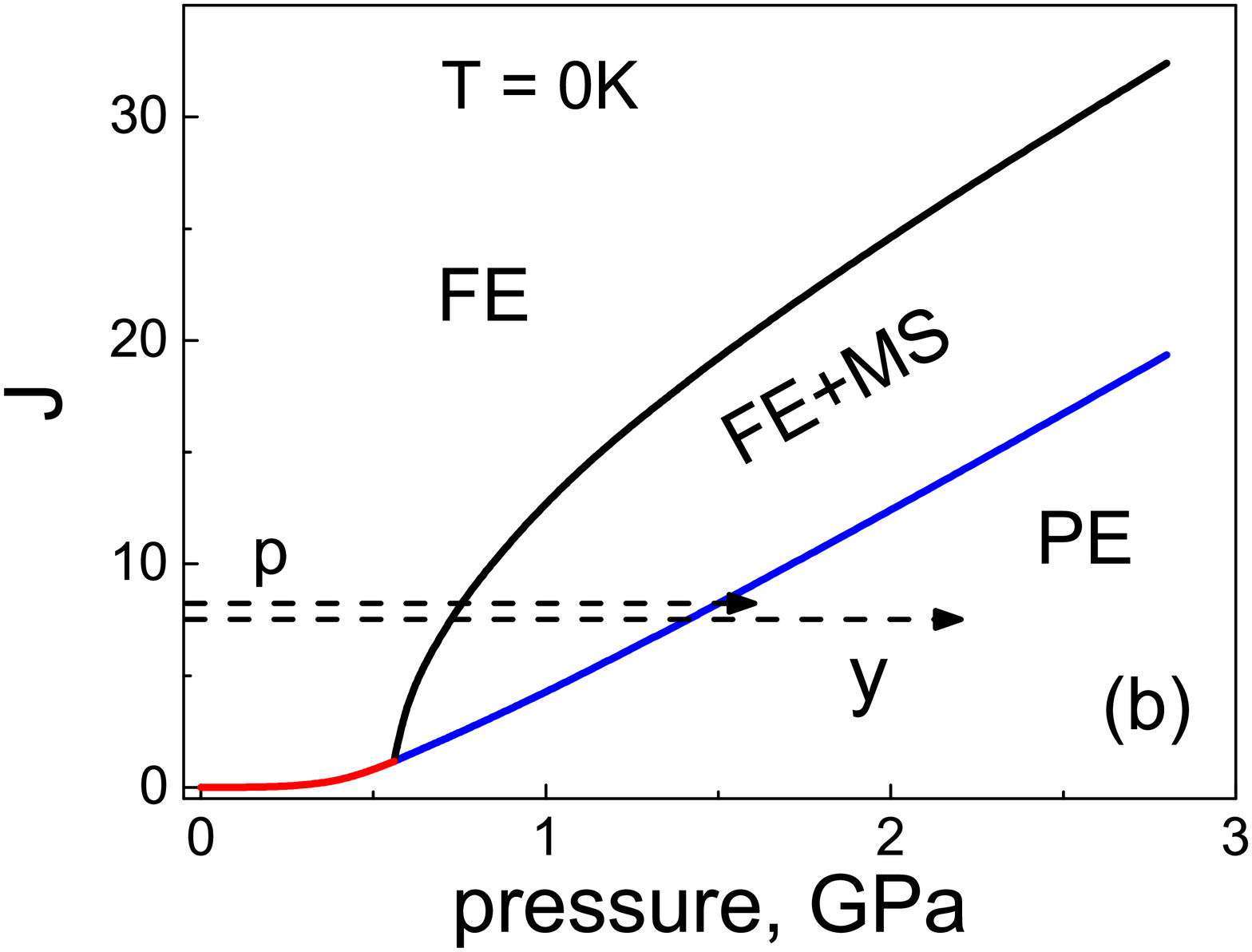}
  \caption{The diagram of states at temperature variation (a) and at $T=0$~K (b) for the AQO model with paraelectric (PE) and ferroelectric (FE) phases, and with intermediate region of coexisting stable ferroelectric phase and metastable (MS) solutions. The thermodynamic ways are shown: for \SPS\ crystal under compression in the range $0-1.5$~GPa and for a rise of lead concentration at normal pressure in \SPbPS\ mixed crystals in the range $0-1$.}\label{fig12}
\end{figure}

Examples of the QAO model solutions are shown at Fig.~\ref{fig11}. Calculated in this model state diagram is shown on Fig.\ref{fig12}. In addition to the paraelectric and ferroelectric phases, it also contain a coexistence region for stable and metastable solutions. This diagram reflects the main features of experimental diagram. For \SPS\ crystal under compression, the second order phase transition temperature lowers from 337~K (at normal pressure) and the tricritical point is reached near 0.6~GPa and at 220~K. The coexistence region of ferroelectric and metastable solutions is delayed till 1.5~GPa. For higher pressure, the paraelectric phase is stabilized. The \PbPS\ compound at normal pressure has quantum paraelectric ground state. By negative pressure below -1.7~GPa the crystal lattice acentricity can be induced.

The dielectric susceptibility temperature dependence in quantum paraelectric \PbPS\ have been calculated in the AQO model and compared with experimental data (Fig.~\ref{fig11}, b). The dielectric susceptibility increase at cooling from room temperature to 20~K obviously reflects strong anharmonicity of the local potential in this compound.

\section{Conclusions}
The valence fluctuations play important role in the nature of ferroelectric and quantum paraelectric states in \SnPb\ semiconductors. The charge disproportionation of phosphorous ions $\rm{P}^{4+} + \rm{P}^{4+}\rightarrow\rm{P}^{3+} + \rm{P}^{5+}$ can be related to recharging of SnPS$_3$ (or PbPS$_3$) structural groups. This approximation permits to consider a simplified model of the crystal lattice as set of the half-filled sites. Experimental temperature-pressure phase diagrams for \SPS\ crystal and temperature-composition one for \SPbPS\ mixed crystals with tricritical point and with decrease of phase transitions lines to 0~K, together with the data about some softening of low energy optic phonons and rise of dielectric susceptibility at cooling in quantum paraelectric state of \PbPS\ crystal, are analyzed by first principles electron and phonon calculations and compared with electronic correlations models. The anharmonic quantum oscillators model is developed for description of temperature-pressure-composition phase diagram shape. Temperature dependence of dielectric susceptibility is also described in this model. The chemical bonds covalence obviously complicates picture of the charge disproportionation which can be presented as $\rm{P}^{4+} + \rm{P}^{4+}\rightarrow\rm{P}^{(4-x)+} + \rm{P}^{(4+x)+}$, and parameter $x$ is interesting for determination at the further development of electronic correlation models.

\section*{Acknowledgments}
Authors acknowledge Prof. Ihor V. Stasyuk for fruitful discussions.

\end{document}